\documentclass[aps,prc,twocolumn,amsmath,showpacs,floatfix]{revtex4}
\usepackage{graphicx}
\usepackage{epsfig}
\begin{document}

\title{Reply to comment ``On the test of the modified BCS at finite 
temperature''}

\author{V.~Yu.~Ponomarev$^{1,2}$ and A.~I.~Vdovin$^{1}$}
\affiliation{
$^1$Bogoliubov Laboratory of Theoretical Physics,
Joint Institute for Nuclear Research,
141980 Dubna, Russia \\
$^2$Institut f\"ur Kernphysik, Technische Universit\"at Darmstadt,
D--64289  Darmstadt, Germany}
\date{\today}


\maketitle

The applicability, predictive power, and internal consistency of 
a modified BCS (MBCS) model suggested by Dang and Arima
have been analyzed in details in \cite{pv}. 
It has been concluded that ``The $T$-range of the MBCS applicability can 
be determined as far below the critical temperature $T_c$'', i.e.,
$T<<T_c$.
Unfortunately, the source of our conclusions has been misrepresented 
in \cite{DA06} and referred to MBCS predictions at $T>>T_c$.

Since above $T_c$ particles and holes contribute to an MBCS gap with 
opposite signs, the model results are rather 
sensitive to details of a single particle spectra (s.p.s.) 
(e.g., discussion in Sec.~IV.~A.~1. of \cite{DA03b}). 
As so, it is indeed possible to find conditions when the MBCS simulates
reasonable thermal behavior of a pairing gap. This can be achieved,
e.g., by introducing some particular $T$-dependence of the s.p.s.
(entry 1 in \cite{DA06}) or by adding an extra level to a picket
fence model (PFM) (entry 2 in \cite{DA06}). But such results are
very unstable and accordingly, the model has no predictive
power.

Dang and Arima explain poor MBCS results for the PFM 
(${\rm N}=\Omega=10$) discussed in \cite{pv} by referring to strong 
asymmetry in the line shape of the quasiparticle-number fluctuations 
$\delta {\cal N}_j$ above $T \sim 1.75$~MeV
(symmetry of $\delta {\cal N}_j$ is announced as a criterion of
the MBCS applicability.)
The space limitation is blamed for that in \cite{DA06}. 
Remember, particle-hole symmetry is an essential feature of the 
PFM with $N=\Omega$. 
Thus, strong asymmetry is reported from the MBCS calculation in an 
ideally symmetric system.

\begin{figure}
\epsfig{figure=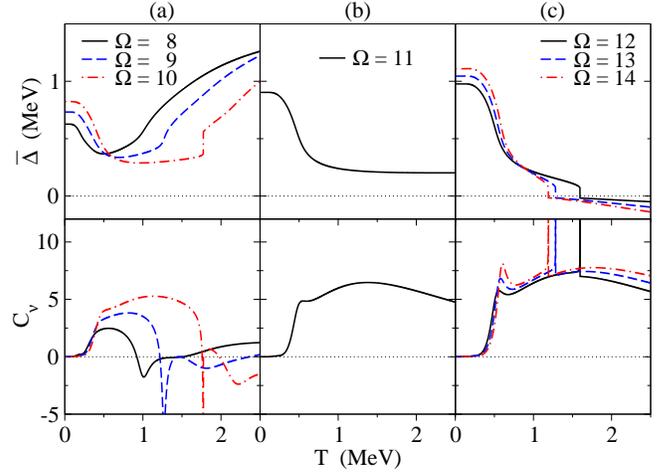,width=85mm,angle=0}
\caption{\label{fig1}
(Color online) The MBCS pairing gap $\bar{\Delta}$ (top panels) and
specific heat $C_{\nu}$ (bottom panels) for the PFM with $N=10$ 
and (a) $\Omega=8,9,10$, (b) $\Omega=11$, and (c) $\Omega=12,13,14$. 
The pairing strength $G = 0.4$~MeV in all cases.
}
\end{figure}

It has been found that a less symmetrical example $N=10,\, \Omega=11$ 
satisfies better the MBCS criterion \cite{DA06}.
Indeed, the model mimics behavior of a macroscopic theory in 
this case (see Fig.~\ref{fig1}b). 
But this example is the only one where the MBCS does not 
break, in a long row of physically very close examples with more limited 
or less limited s.p.s..
In all other examples we witness either negative heat capacity $C_{\nu}$
(Fig.~\ref{fig1}a) or negative gap $\bar{\Delta}$ 
(Fig.~\ref{fig1}c) at rather moderate $T$ (see also \cite{Civ89}).

Unfortunately, conclusion in \cite{DA06} that ``within extended 
configuration spaces $\ldots$ the MBCS is a good approximation up to 
high $T$ even for a system with $N=10$ particles'', is based on a 
single example while in all other $N=10$ examples the MBCS 
yields unphysical predictions.

The most serious problem of the MBCS is its thermodynamical inconsistency.
It is not sufficient to declare two quantities $<H> = {\rm Tr}(HD)$ and 
$\cal{E}$ representing the system energy, analytically equal by definition 
(as is done in footnote [8] of \cite{DA06}) to prove the model consistency.
It is easy to find that expression for $\cal{E}_{\rm MBCS}$ 
(in the form of Eq.~(83) in \cite{DA03b}) can be obtained 
in the same way as all other MBCS equations have been derived:
straightforwardly replacing the Bogoliubov $\{u_j,\ v_j\}$ coefficients
in $\cal{E}_{\rm BCS({\it T}=0)}$ expression  
by $\{\bar{u}_j,\ \bar{v}_j\}$ coefficients.
Numeric results in Fig.~9 of \cite{pv} show that 
$<H>_{\rm MBCS}$ and $\cal{E}_{\rm MBCS}$ have nothing in common
while  $<H>_{\rm BCS} \approx \cal{E}_{\rm BCS}$
as it should be for thermodynamically consistent theory.

\begin{figure}
\epsfig{figure=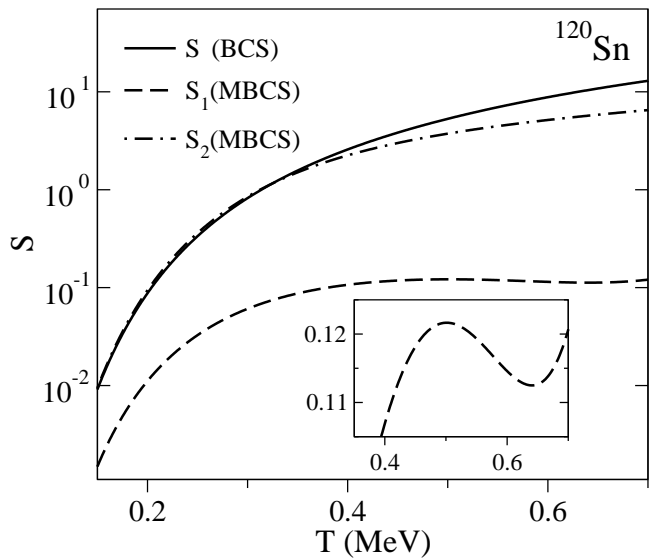,width=85mm,angle=0}
\caption{\label{fig2}
Entropy of the neutron system in $^{120}$Sn calculated within
the FT-BCS (solid curve) and MBCS (dashed and dot-dashed curves).
Notice the logarithmic $y$ scale of the main figure and linear $y$ scale 
of the insert.
See text for details.}
\end{figure}

Another example of the MBCS thermodynamical inconsistency is shown below.
We calculate the system entropy $S$ as
\begin{equation}
S_1 = \int_0^T \, \frac{1}{t} \,\cdot \,
\frac{\partial \cal{E}}{\partial t} \,dt
\nonumber
\end{equation}
and
\begin{equation}
S_2 = - \sum_j \, (2j+1)\, \left [ \, n_j \, {\rm ln} \, n_j + (1-n_j) \, 
{\rm ln} \, (1-n_j) \right ] 
\nonumber
\end{equation}
where $n_j$ are thermal quasiparticle occupation numbers.
In Fig.~\ref{fig2} we compare $S_1$ and $S_2$ quantities which refer to 
thermodynamical and statistical mechanical definition of entropy, 
respectively. 
The calculations have been performed for neutron system of $^{120}$Sn
with a realistic s.p.s.

It is not possible to distinguish by eye $S_1$ and $S_2$ in the FT-BCS 
calculation (solid curve in Fig.~\ref{fig2} represents both quantities) 
as it should be for thermodynamically consistent theory.
The MBCS $S_1$ and $S_2$ quantities are shown by dashed and dot-dashed
lines, respectively.
They are different by orders of magnitude in the MBCS prediction.

We stress that low $T$ part is presented in Fig.~\ref{fig2}.
Dramatic disagreement between $S_1({\rm MBCS})$ and $S_2({\rm MBCS})$
representing the system entropy remains at
higher $T$ as well but we do not find it necessary to extend the
plot: the model obviously does not describe correctly a heated system 
even at $T \sim 200$~keV. 

We show in the insert of Fig.~\ref{fig2} another MBCS prediction: 
entropy $S_1$ decreases as temperature increases.
This result is very stable against variation of the pairing strength
$G$ within a wide range and contradicts the second law of 
thermodynamics.

Finally, we repeat, the conclusion in \cite{pv} that ``The $T$-range 
of the MBCS applicability can be determined as far below the critical 
temperature $T_c$'' is based on the analysis of the model predictions 
from $T<<T_c$ and not on $T>>T_c$ results as presented in \cite{DA06}.

The work was partially supported by the Deutsche Forschungsgemeinschaft 
(SFB 634).

\end{document}